\documentclass[aps,prl,floatfix,epsfig,twocolumn,showpacs,preprintnumbers]{revtex4}
\usepackage{amssymb}
\usepackage{amsmath}
\usepackage{graphicx}
\usepackage{epstopdf}
\usepackage{dcolumn}
\usepackage{bm}
\usepackage{mathrsfs}
\usepackage[pdfstartview=FitH, CJKbookmarks=true, bookmarksnumbered=true, bookmarksopen=true, pdfborderstyle={/S/U}, pdfborder={0 0 1}, colorlinks=true, linkcolor=blue, anchorcolor=green, citecolor=blue]{hyperref}

\DeclareMathAlphabet      {\mathbf}{OT1}{cmr}{bx}{n}
\begin{document}

\title{Competing d$_{xy}$ and s$_{\pm }$ Pairing Symmetries in
Superconducting La$_{3}$Ni$_{2}$O$_{7}$ emerge from LDA+FLEX Calculations}
\author{Griffin Heier$^{1}$, Kyungwha Park$^{2}$, Sergey Y. Savrasov$^{1}$}
\affiliation{$^{1}$Department of Physics, University of California, Davis, CA 95616, USA}
\affiliation{$^{2}$Department of Physics, Virginia Tech, Blacksburg, Virginia 24061, USA}

\begin{abstract}
With recent discoveries of superconductivity in infinite--layer nickelates,
and in La$_{3}$Ni$_{2}$O$_{7}$ under high pressure, new opportunities
appeared that yet another family of high--temperature superconductors based
on Ni element may exist in Nature as was previously the case of cuprates and
iron based materials. With their famous strong Coulomb correlations among 3d
electrons and the proximity to antiferromagnetic instability these systems
represent a challenge for their theoretical description, and most previous
studies of superconductivity relied on the solutions of simplified
few--orbital model Hamiltonians. Here, on the other hand, we use a recently
developed combination of density functional theory with momentum and
frequency resolved self--energies deduced from the so--called
Fluctuational--Exchange (FLEX)--type Random Phase Approximation (RPA) to
study spin fluctuation mediated pairing tendencies in La$_{3}$Ni$_{2}$O$_{7}$
under pressure. This methodology uses first--principle electronic structures
of an actual material and is free of tight--binding parametrizations
employed in model Hamiltonian approach. Based on our numerical
diagonalization of the BCS\ Gap equation we show that competing d$_{xy}$ and
s$_{\pm }$ pairing symmetries emerge in superconducting La$_{3}$Ni$_{2}$O$%
_{7}$ with the corresponding coupling constants becoming large in the
proximity of spin density wave instability. The results presented here are
discussed in light of numerous other calculations and provide on--going
experimental efforts with predictions that will allow further tests of our
understanding of unconventional superconductors.
\end{abstract}

\date{\today }
\maketitle

\section{I. Introduction}

Opportunities to realize high--temperature superconductivity have always
been a subject of enormous research interest, and recent discoveries of
superconducting nickelates\cite{Nature-2019,Nature-La3Ni2O7} is not an
exception. Despite numerous theoretical and experimental efforts in the past 
\cite%
{LaNiO2-exp-JPCS-1996,LaNiO2-exp-JAC-1998,LaNiO2-exp-JACS-1999,LaNiO2-exp-Phys.C.-2013,Pickett-2004,LaNiO2-LDA+U-1999}%
, infinite--layer Nd$_{2}$NiO$_{2}$ were shown to exhibit superconductivity
at 8K only a few years ago \cite%
{Nature-2019,Superconducting-Dome-1,Superconducting-Dome-2,PrNiO2} but just
discovered 80K superconductivity in bulk La$_{3}$Ni$_{2}$O$_{7}$ under
applied pressure over 14GPa\cite{Nature-La3Ni2O7} has brought nickelates
into focus of becoming a new addition to the famous family of
high--temperature superconducting cuprates and ironates\cite%
{HTC-1,HTC-2,HTC-3}.

This breakthrough has inspired\ a large--scale theoretical effort to
understand the nature of superconductivity in La$_{3}$Ni$_{2}$O$_{7}$. First
principles electronic structure calculations based on Density Functional
Theory (DFT) and Local Density Approximation (LDA)\cite{DFT} reveal the
dominant role of Ni 3d$_{x^{2}-y^{2}}$ orbitals and, in addition, of Ni 3d$%
_{3z^{2}-r^{2}}$ orbitals \cite{Nature-La3Ni2O7,Pickett-2011,LDA+U-arxiv}.
Correlation effects beyond DFT have been studied \cite%
{Werner-2023,Leonov-2023,Lechermann-2023} using combinations of LDA and GW
methods with dynamical mean field theory (LDA+DMFT and GW+DMFT)\cite%
{DFT+DMFT} . A two--orbital bilayer model for the bands near the Fermi level
has been proposed to describe the low energy physics of this material \cite%
{Bilayer1,Bilayer2}. Many--body simulations of model Hamiltonians with
Quantum Monte Carlo\cite{QMC-arxiv} and density matrix renormalization group%
\cite{DMRG-arxiv} techniques have also recently appeared.

Focusing on the superconducting state, the discussion of the pairing
symmetry has been the subject of extensive studies in a series of recent
works\cite%
{Hu-arxiv,Qin-2023,Yang-2023,Tian-2023,Dagotto-arxiv,Lechermann-2023,SPM-arxiv,LongRange}%
. Within the two--orbital tight--binding model, the pairing instability was
studied using functional renormalization group (FRG) approach and the
multi--orbital $t-J$ model. An s$_{\pm }$--wave pairing with sign--reversal
gaps on different Fermi surfaces is revealed, reminiscent of iron based
superconductors \cite{Hu-arxiv}. Employing the static auxiliary field Monte
Carlo approach to simulate a minimal effective model containing local d$%
_{3z^{2}-r^{2}}$ interlayer spin singlets and metallic d$_{x^{2}-y^{2}}$
bands, the authors of Ref. \cite{Qin-2023} have reached a similar
conclusion, together with another FRG calculations that also yielded s$_{\pm
}$--wave Cooper pairing triggered by the spin fluctuations\cite{Yang-2023}.
The cellular version of the DMFT was used on the two--orbital Hubbard model
where the s$_{\pm }$--superconductivity was observed\cite{Tian-2023}.

Several calculations of the pairing interaction have been performed\ \cite%
{Lechermann-2023,Dagotto-arxiv,SPM-arxiv,LongRange} using the two--orbital
model and the Random Phase Approximation (RPA) -- the method first used to
understand properties of heavy fermion systems many years ago \cite%
{Scalapino,Varma}. Diagrammatically, the RPA includes particle--hole ladder
and bubble diagrams\cite{Paramagnons,BerkSchrieffer}. The inclusion to these
series of particle--particle ladder diagrams has also been proposed in the
past, which was entitled as the Fluctuational Exchange Approximation (FLEX) 
\cite{FLEX}. This contribution was however found to be not essential in the
proximity to magnetic instability where the most divergent terms are
described by the particle--hole ladders\cite{Muller}.

Using the RPA, the intra--orbital Coulomb interaction parameter $U=0.8$ eV, $%
J=U/4$ and the inter--orbital repulsion $V=U-2J$, the pairing instability
was shown to be induced in the s$_{\pm }$--wave channel due to nesting
between the ($1/2,1/2$)2$\pi /a$ and ($1/2$, 0)2$\pi /a$ points in the
Brillouin zone \cite{Dagotto-arxiv}. A different RPA calculation \cite%
{Lechermann-2023} discussed the appearance of multiple leading symmetries (d$%
_{x^{2}-y^{2}}$ vs. d$_{xy}\ $) in the calculation with $U=0.36$ eV and by
utilizing either $J=U/4$ or $U/7,$ while the solution of the s$_{\pm }$
symmetry was found as the sub--leading one. The study of the paring symmetry
as a function of $U$ and using the Kanamori rule: $J=U/4,$ $V=U-2J,$ has
been performed in Ref. \cite{SPM-arxiv}, where it was shown that the s$_{\pm
}$ slightly dominates over the d$_{xy}$ symmetry once the system is tuned to
the spin--density--wave (SDW) instability occurring for the two--orbital
model at $U$ slightly above 1.2eV. A most recent work \cite{LongRange} has
utilized maximally--localized--Wannier--function construction and found that
the superconducting symmetry of La$_{3}$Ni$_{2}$O$_{7}$ is robustly d$_{xy}$
if its LDA band structure is accurately reproduced in the downfolded model.

We have recently implemented a combination of density functional electronic
structure theory with momentum-- and energy--dependent self--energy deduced
from the above mentioned FLEX--RPA approach\cite{LDA+FLEX}. Such method is
free of tight--binding parametrizations and utilizes full electronic energy
bands and the wave functions available in the density functional
calculation. It\ evaluates dynamical charge and spin susceptibilities of the
electrons in a Hilbert space restricted by correlated orbitals only, similar
to popular LDA+U\cite{DFT+U} and LDA+DMFT approaches\cite{DFT+DMFT}.
Evaluations of superconducting pairing interactions describing scattering of
the Cooper pairs at the Fermi surface in a realistic material framework
became possible using this method. Our most recent calculations of HgBa$_{2}$%
CuO$_{4}$\cite{Hg-FLEX}, a prototype single--layer cuprate superconductor,
where a much celebrated d$_{x^{2}-y^{2}}$ symmetry of the order parameter
was easily recovered with LDA+FLEX, has demonstrated applicability of this
method to study spin--fluctuation mediated superconductivity without
reliance on tight--binding approximations of their electronic structures.

\begin{figure}[tbp]
\includegraphics[height=0.366\textwidth,width=0.40%
\textwidth]{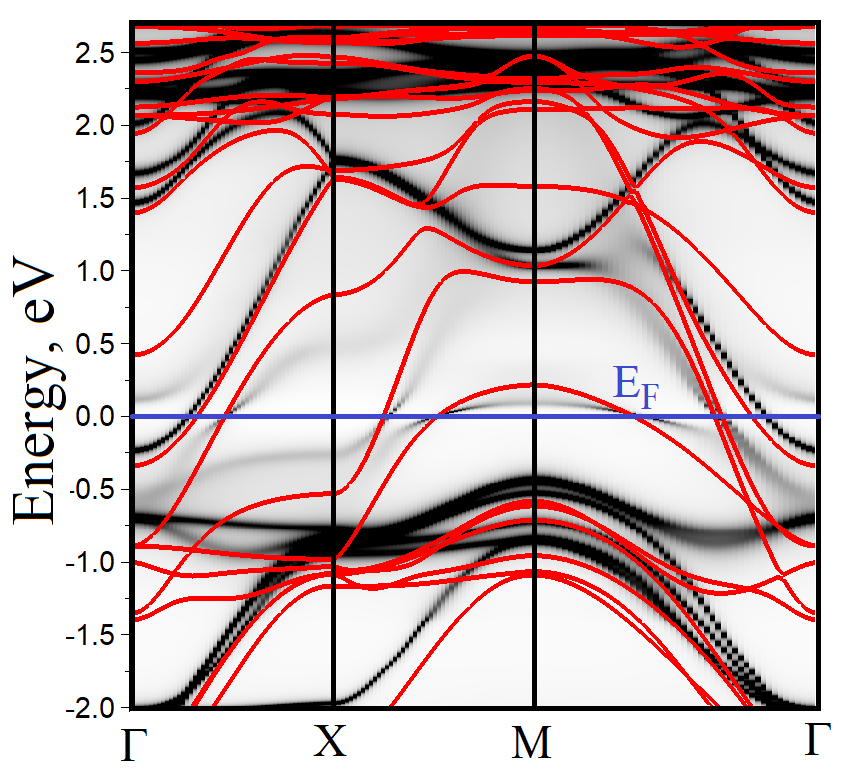}
\caption{Electronic structure of La$_{3}$Ni$_{2}$O$_{7}$ with experimentally
determined lattice parameters that correspond to pressure 30 GPa calculated
using LDA+FLEX method (gray shading) with U=3 eV and J=0.5eV, as well as the
result of the density functional LDA calculation (red lines).}
\label{FigBands}
\end{figure}

Here we apply the LDA+FLEX method to study the superconductivity in bulk La$%
_{3}$Ni$_{2}$O$_{7}$ under pressure$.$ Our numerically evaluated
superconducting pairing interaction describing scattering of the Cooper
pairs at the Fermi surface is used to exactly diagonalize the linearized
Bardeen--Cooper--Schrieffer (BCS) gap equation on a three dimensional
k--grid of points in the\ Brillouin Zone. The highest eigenstate $\lambda
_{\max }$ deduced from this procedure represents a spin fluctuational
coupling constant similar to electron--phonon $\lambda _{e-p}$ in
conventional theory of superconductivity. The spin fluctuational $\lambda
_{\max }$ was found to be very sensitive to the actual values of the Hubbard
interaction $U$ among Ni 3d electrons serving as input to this calculation
but reaches rather large values in close proximity to the spin density wave
instability. This is in accord with recent RPA study of this system using
the tight--binding simulation \cite{SPM-arxiv}.

The same sensitivity to input $U$ is seen in our calculated normal state
self--energies which were found to show a weak $\mathbf{k-}$ and strong
frequency dependence with particularly large electronic mass renormalization 
$m^{\ast }/m_{LDA}=1+\lambda _{sf}$ $\ $in the proximity to SDW. Both $%
\lambda _{\max }$ and $\lambda _{sf\text{ }}$ determine the renormalized
coupling constant $\lambda _{eff}=\lambda _{\max \text{ }}/(1+\lambda _{sf})$
that enters the BCS expression for $T_{c}\symbol{126}\omega _{sf}\exp
(-1/\lambda _{eff}),$ where $\omega _{sf}$ is the characteristic energy of
spin fluctuations. We find $\lambda _{eff}$ to be modest and incapable to
deliver high critical temperatures unless we tune $U$ to be very close to
the SDW point. Two competing symmetries of the superconducting order
parameter, d$_{xy}$ vs s$_{\pm },$ are seen from the analysis of the highest
eigenstates of the BCS\ gap equation. This result is also found in our own
and published \cite{Lechermann-2023,Dagotto-arxiv,SPM-arxiv,LongRange}
tight--binding RPA simulations . The realization of d$_{xy}$ would be quite
unusual in light of d$_{x^{2}-y^{2}}$ pairing state being found in cuprates%
\cite{RMP-1995} and of s$_{\pm }$ pairing in ironates. Due to the appearance
of nodes in d$_{xy}$ and its absence in s$_{\pm }$ symmetry the on--going
experimental efforts should easily elucidate which pairing symmetry is
realized here.

Our paper is organized as follows: In Section II we discuss the results of
the correlated electronic structure for La$_{3}$Ni$_{2}$O$_{7}$ using the
LDA+FLEX formalism. In Section III we present our results of exact
diagonalization of the linearized BCS equation and correspondingly extracted
superconducting energy gaps and the eigenvalues as a function of $U$. We
also give estimates of the spin fluctuational mass enhancement that
determines the effective coupling constant $\lambda _{eff}$ entering the BCS
expression for the $T_{c}$. In Section IV, we repeat the calculation using
the two--orbital bilayer tight--binding model and give comparisons with our
full LDA+FLEX calculation. Section V is the conclusion.

\begin{figure}[tbp]
\includegraphics[height=0.720\textwidth,width=0.40%
\textwidth]{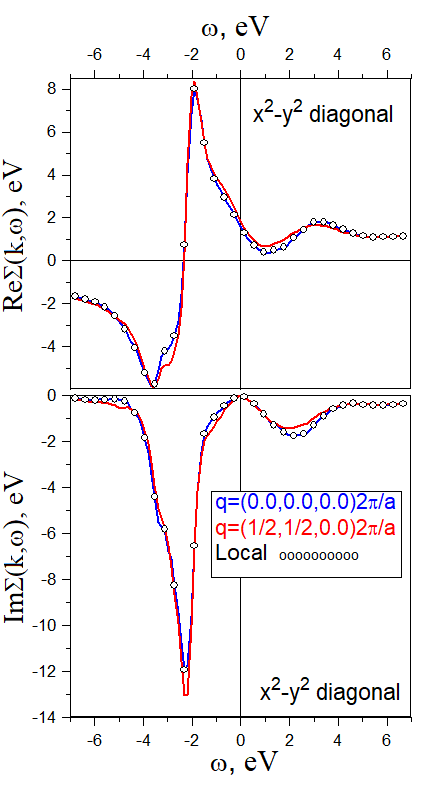}
\caption{{}Calculated self--energy $\Sigma (\mathbf{k},\protect\omega )$
(top is the real part, and bottom is imaginary part) using FLEX--RPA
approximation for d$_{x^{2}-y^{2}}$ electrons of Ni in La$_{3}$Ni$_{2}$O$%
_{7} $. The wave vectors k correspond to ($000$) and ($\frac{1}{2},\frac{1}{2%
},0)$ of the Brillouin Zone. The circles show the result of the local
self--energy approximation taken as the average over all k--points. A
representative value of Hubbard $U$=3 eV and J=0.5eV is used.}
\label{FigSigma}
\end{figure}

\section{II. Electronic Structure of La$_{3}$Ni$_{2}$O$_{7}$ from LDA+FLEX}

A unit cell of bulk La$_{3}$Ni$_{2}$O$_{7}$ contains two Ni--O planes and is
expected to have its orthorhombic structure with space group $Fmmm$ \cite%
{Greenblatt}. This is also true under applied pressure\cite{Nature-La3Ni2O7}%
. At 30 GPa, the theoretically deduced lattice parameters are given by: $%
a=5.29\mathring{A},b=5.21\mathring{A},c=19.73\mathring{A}$. We perform our
density functional electronic structure calculations \cite{DFT} using the
full potential linear muffin--tin orbital method \cite{FPLMTO}. The
self--energies for Ni 3d electrons, $\Sigma (\mathbf{k},\omega ),$ are
evaluated on the 12x12x12 grid of $\mathbf{k}$--points and at the frequency
range between -13.6 eV and +13.6 eV from the Fermi energy based on RPA--FLEX
procedure described previously\cite{LDA+FLEX,Hg-FLEX}. We use the values of
an on--site Hubbard interaction $U=3eV$ and $J=0.5eV$ as an input to the
simulation similarly to Ref. \cite{Bilayer1}.

Since the self--energy has both real and imaginary parts, the electronic
states do no longer have infinite life times. We evaluate the poles of the
single particle Green function, and the obtained $\text{Im}G(\mathbf{k}%
,\omega )$ for La$_{3}$Ni$_{2}$O$_{7}$ is shown in Fig. \ref{FigBands}. The
Fermi surface states mainly consist of Ni--$3d_{_{x^{2}-y^{2}}}$ and Ni--$%
3d_{3z^{2}-r^{2}\text{ }}$orbital character. There is a small La based
pocket around the $\Gamma $ point whose exact position depends on whether
one uses theoretical (as in some previous works\cite{Bilayer1}) or
experimental atomic positions utilized here. Such sensitivity was
investigated in details in Ref. \cite{Werner-2023}, but, in any case it
should be irrelevant for the superconducting behavior of La$_{3}$Ni$_{2}$O$%
_{7}$.

Most of the poles are seen as sharp resonances (plotted in black) in the
function $\text{Im}G(\mathbf{k},\omega )$ that closely follows the LDA
energy dispersions (plotted in red). However, the difference is seen in the
behavior of the Ni 3d$_{_{x^{2}-y^{2}}}$ and Ni-3d$_{3z^{2}-r^{2}\text{ }}$%
states in the vicinity of the Fermi surface that acquire a strong damping at
energies away from the Fermi level. As the primary effect of the
self--energy is the renormalization of the electronic bandwidth, we found
the electronic mass enhancement to be around 3.4 for Ni 3d$_{_{x^{2}-y^{2}}}$
and and 4.7 for Ni 3d$_{3z^{2}-r^{2}\text{ }}$ orbitals using the setup with 
$U=3eV$ and $J=0.5eV$, but it also needs to be noticed that the mass
enhancement depends strongly on $U$ as we discuss later in this work.

We further plot the diagonal elements of the real and imaginary parts of the
self--energy for Ni $3d_{_{x^{2}-y^{2}}}$ orbital in Fig.\ref{FigSigma}. To
illustrate the $\mathbf{k}$--dependence we show the result for the two
points of the Brillouin Zone (BZ): $\Gamma =(0,0,0)$ (blue) and $%
M=(1/2,1/2,0)$ (red)$.$ We generally find the k--dependence of $\Sigma (%
\mathbf{k},\omega )$ to be quite small prompting the locality feature of the
self--energy. Very similar behavior is seen for other $\mathbf{k}$--points
of the BZ and also for the matrix elements corresponding to Ni$%
_{d_{3z^{2}-r^{2}}}$ orbitals. The local self--energy $\Sigma _{loc}(\omega
) $ can be evaluated\ as an integral over all $\mathbf{k}$--points, and its
frequency dependence is shown in Fig. \ref{FigSigma} by small circles. We
conclude that there is a close agreement between $\Sigma _{loc}(\omega )$
and $\Sigma (\mathbf{k},\omega )$.

A pole--like behavior for the self--energy at frequencies around 2 eV is
also seen in our LDA+FLEX calculation. Those poles are led to additional
resonances in the single--particle Green functions that cannot be obtained
using static mean DFT\ based approaches. The imaginary part of the
self--energy nearly diverges indicating strongly damped excitations. Those
resonances are usually hard to associate with actual energy bands.

\begin{figure}[tbp]
\includegraphics[height=0.786\textwidth,width=0.40%
\textwidth]{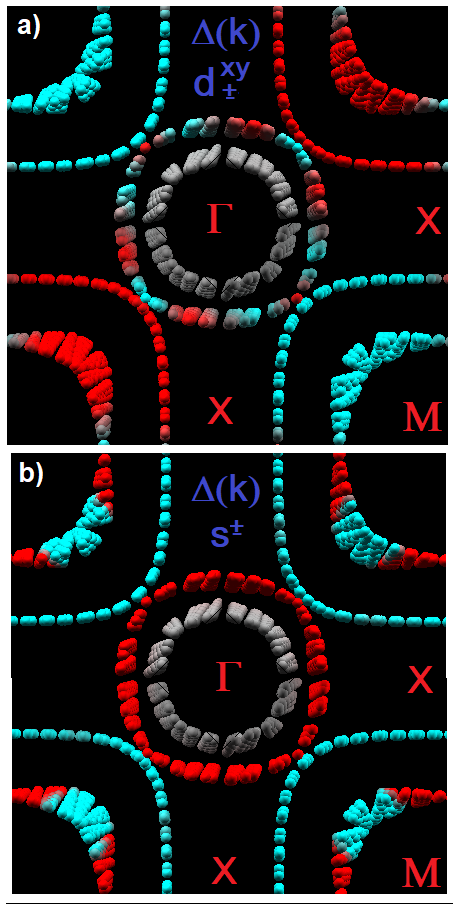}
\caption{{}Calculated superconducting energy gap $\Delta (\mathbf{k})$ for
singlet pairing in La$_{3}$Ni$_{2}$O$_{7}$ using numerical solution of the
linearized BCS gap equation with the pairing interaction evaluated using
LDA+FLEX(RPA)\ approach. Blue/red color marks points at the quasi 2D Fermi
surface (viewed from the top) that corresponds to the negative/positive
values of $\Delta (\mathbf{k}),$ while the values around zero are colored in
gray. Case a) is the eigenstate with d$_{xy}$ symmetry and b) is the
eigenstate with s$_{\pm }$ symmetry.}
\label{FigGaps}
\end{figure}

\section{III. Superconducting Properties of La$_{3}$Ni$_{2}$O$_{7}$ from
LDA+FLEX}

We utilize our LDA+FLEX(RPA) method to evaluate the spin fluctuation
mediated pairing interaction. The Fermi surface is triangularized onto small
areas described by about 6,000 Fermi surface momenta for which the matrix
elements of scattering between the Cooper pairs are calculated using the
approach described in Ref. \cite{Hg-FLEX}. The\ linearized BCS\ gap equation
is then exactly diagonalized and the set of eigenstates is obtained for both
singlet ($S=0$) and triplet ($S=1$) Cooper pairs. The highest eigenvalue $%
\lambda _{\max }$ represents the physical solution and the eigenvector
corresponds to superconducting energy gap $\Delta _{S}(\mathbf{k}j)$ where $%
\mathbf{k}$ is the Fermi surface momentum and $j$ numerates the Fermi
surface sheets.

\begin{figure}[tbp]
\includegraphics[height=0.329\textwidth,width=0.40%
\textwidth]{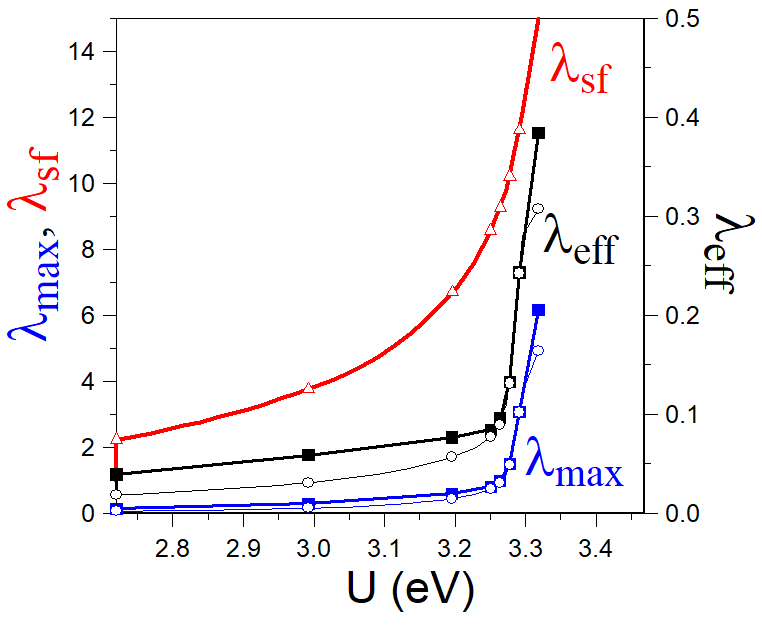}
\caption{{}Calculated using LDA+FLEX(RPA)\ method dependence of spin
fluctuational mass enhancement, $\protect\lambda _{sf},$ and the two highest
eigenvalues $\protect\lambda _{\max }$ (squares and circles) corresponding
to d$_{xy}$ and s$_{\pm }$ symmetries of the linearized BCS equation as a
function of the on--site Hubbard interaction U for d-electrons of Ni in La$%
_{3}$Ni$_{2}$O$_{7}$. The effective coupling constant $\protect\lambda %
_{eff}=\protect\lambda _{\max }/(1+\protect\lambda _{sf})$ is shown on the
right scale. }
\label{FigLambdas}
\end{figure}

During the course of the diagonalization we however find that there are two
highest eigenvalues that appear very close to each other. We analyze the
behavior of $\Delta _{S}(\mathbf{k}j)$ corresponding to them as a function
of the Fermi momentum using the values of $U=3eV$ and $J=0.5eV$. They are
both related to the spin singlet states, and Fig.\ref{FigGaps} a) and b)
shows first and second highest eigenstates, respectively. One can see that
the most favorable eigenstate $\Delta _{S=0}(\mathbf{k}j)$ shows the
behavior of a d--wave with $xy$ symmetry (zeroes pointing along $k_{x}$ and $%
k_{y}$ directions) The plot distinguishes negative/positive values of $%
\Delta $ by blue/red colors while zeroes of the gap function are colored in
grey\ as, for example, the case of the La pocket seen around $\Gamma $
point. This result is interesting because it differs from the predicted
behavior in cuprates with their d$_{x^{2}-y^{2}}$ symmetry and also from
iron pnictides with their s$_{\pm }$ behavior. As emphasized earlier \cite%
{Dagotto-arxiv}, the Fermi surface nesting here is quite strong for both BZ
edges at (1/2,0,0)2$\pi /a$ and (1/2,1/2,0)2$\pi /a$\ \ which results in a
delicate competition between the two spin density wave instabilities.

Fig.\ref{FigGaps} (b) shows the behavior of the second highest eigenstate.
Here we clearly resolve sign changing energy gap values for different sheets
of the Fermi surface corresponding to the s$_{\pm }$ symmetry. This behavior
was recently discussed in several model calculations\cite%
{Hu-arxiv,Qin-2023,Yang-2023,Tian-2023}. It was also found explicitly using
the RPA calculations with the two--orbital bilayer model \cite%
{Dagotto-arxiv,SPM-arxiv}.

\begin{figure}[tbp]
\includegraphics[height=0.404\textwidth,width=0.40\textwidth]{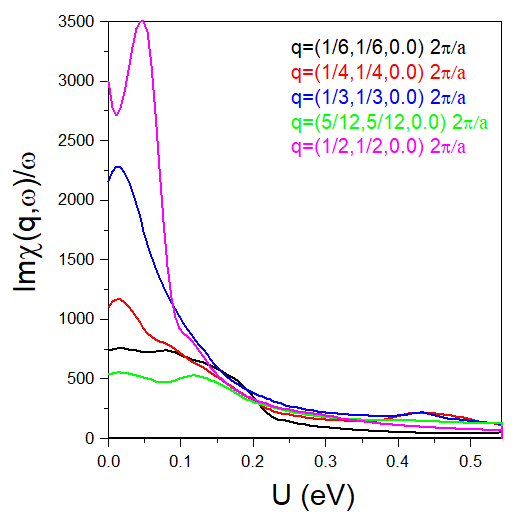}
\caption{{}Calculated imaginary part of spin susceptibility weighted by $%
\protect\omega ^{-1}$ in La$_{3}$Ni$_{2}$O$_{7}$ plotted as a function of
frequency for several wave vectors $q$ traversing from the $\Gamma $ to $M$
point of the BZ and for the Ni d$_{3z^{2}-r^{2}}$ diagonal elements of the
spin susceptibility.}
\label{FigImHi}
\end{figure}

Several works have emphasized the sensitivity of the solution to both the
input interaction parameterss $U$\cite{SPM-arxiv}, $J$\cite{Lechermann-2023}
and the longer range hopping integrals \cite{LongRange} resulting in the
appearance of the competing symmetries, such as s$_{\pm },$d$_{xy}$ as well
as d$_{x^{2}-y^{2}}$. Since we use full LDA\ derived band structures in the
simulation, we can gain additional insight by varying the parameter $U.$ We
use its range of values between the 2.8 and 3.3 eV and extract from the BCS\
gap equation the highest eigenstates and their symmetries as a function of $U
$. This plot is shown in Fig.\ref{FigLambdas} where one can see the
dependence of the highest (squares) and the next highest (circles)
eigenvalues as a function of $U.$ We find that d$_{xy}$ symmetry dominates
slightly in the range of $U^{\prime }$s below 3.2 eV but competes closely
with the $s_{\pm }$ symmetry once we approach the spin density wave
instability that occurs above 3.2 eV. At the lack of rigorous procedure for
determining $U,$ it is clear that the leading paring symmetry in La$_{3}$Ni$%
_{2}$O$_{7}$ cannot be exactly predicted using this method and the question
should be settled by the experiment.

To get estimates of the critical temperature, we recall that it is not the
eigenvalue $\lambda _{\max }$ but an effective coupling constant $\lambda
_{eff}$ that enters the BCS $T_{c}$ expression: $T_{c}\approx \omega
_{sf}\exp (-1/\lambda _{eff}).$ It incorporates the effects associated with
the mass renormalization describing by the parameter $\lambda _{sf},$ and is
also weakened slightly by the Coulomb pseudopotential $\mu _{m}^{\ast }$
which should refer to the same pairing symmetry $m$ as $\lambda _{\max }$: 
\begin{equation}
\lambda _{eff}=\frac{\lambda _{\max }-\mu _{m}^{\ast }}{1+\lambda _{sf}}
\label{Leff}
\end{equation}

The mass enhancement can be evaluated as the Fermi surface (FS) average of
the electronic self--energy derivative taken at the Fermi level 
\begin{equation}
\lambda _{sf}=-\langle \frac{\partial \Sigma (\mathbf{k},\omega )}{\partial
\omega }|_{\omega =0}\rangle _{FS}  \label{Lsf}
\end{equation}%
We calculate the dependence of $\lambda _{sf}$ on $U$ using analytical
differentiation of the self--energy at zero frequency by utilizing its
spectral representation \cite{Winter}. $\lambda _{sf}$ is found to grow
rapidly in the vicinity of SDW as we illustrate it in Fig.\ref{FigLambdas}.

To give estimates for the effective coupling constant, $\lambda _{eff},$ we
notice that $\mu _{m}^{\ast }$ is expected to be very small for the pairing
symmetries different from the standard s--wave \cite{Alexandrov}. We
therefore set this parameter to zero. The plot of $\lambda _{eff}=\lambda
_{\max }/(1+\lambda _{sf})$ vs. $U$\ is shown in Fig \ref{FigLambdas} with
its scale given on the right. One can see that the range of these values is
quite modest (\symbol{126}0.1) for $U$'s less than 3.2 eV as compared to
both $\lambda _{\max }$ and $\lambda _{sf},$ primarily due to the fact that
the rise in the eigenvalue of the gap equation, is compensated by the
renormalization effect of the electronic self--energy.

Only in the close proximity to the SDW instability, $\lambda _{eff}$ raises
to the values \symbol{126}0.4. (It can go even further up upon tuning $U$).
However, this corresponds to very large values of $\lambda _{sf}\symbol{126}%
15$, which are likely not very realistic. Although, experimental
determination of the electronic mass enhancement for La$_{3}$Ni$_{2}$O$_{7}$
is not presently available, one can quote corresponding range of values for
high--$T_{c}$ cuprates: ARPES studies of Bi$_{2}$Sr$_{2}$CaCu$_{2}$O$%
_{8+\delta }$ produced 0.5$\lesssim \lambda _{sf}\lesssim $1.7 \cite%
{ARPES-MASS}$.$ A different work\cite{ARPES-MASS2} for Bi$_{2}$Sr$_{2}$CaCu$%
_{2}$O$_{8}$ and also for La$_{2-x}$Ba$_{x}$CuO$_{4}$ reported the estimate $%
1\lesssim \lambda _{sf}\lesssim 2.$ Somewhat larger values of the
self--energy slope, $4\div 8,$ taken for several Fermi momenta have been
seen in ARPES analysis of Bi$_{1.74}$Pb$_{0.38}$Sr$_{1.88}$CuO$_{6+\delta }$%
\cite{ARPES-MASS3}. The value of 2.7 along the nodal line was quoted for YBa$%
_{2}$Cu$_{3}$O$_{6.6}$\cite{ARPES-MASS4}. Quantum oscillations reported $%
m^{\ast }$ range from 1.9 to 5 (in units of the free electron mass) for
various cuprates including the value of $2.45\pm 0.15$ for HgBa$_{2}$CuO$%
_{4+\delta }$\cite{MASS5}.

To get estimates for the range of spin fluctuational energies $\omega _{sf},$
we analyze the behavior of the spin susceptibility that is responsible for
the spin fluctuational pairing. Since the BCS approximation assumes that the
superconducting pairing $K$ operates for the electrons residing at the Fermi
surface only, it can be considered as the static ($\omega =0)$ value for the
dynamically resolved interaction: $K(\mathbf{q},\omega )=I+I\chi (\mathbf{q}%
,\omega )I.$ Here $I$ is the static on--site bare interaction matrix that
incorporates $U$ and $J\ $while $\chi (\mathbf{q},\omega )$ is the
interacting susceptibility matrix. As a result of the Kramers--Kroenig
transformation, Re$K(\mathbf{q},0)$ can be expressed via the inverse
frequency moment of its imaginary part, Im$K(\mathbf{q},\omega )/\omega ,$
that is further proportional to the imaginary part of the spin
susceptibility weighted by $\omega ^{-1}.$ Thus, the frequency resolution of
the spin fluctuational spectrum can be easily analyzed by plotting the
function Im$\chi (\mathbf{q},\omega )/\omega .$ \ We present this data in
Fig.\ref{FigImHi}. The spin susceptibility is the matrix which has four
orbitals corresponding to various d--electron states and also two sites
corresponding to two NiO$_{2}$ planes within the unit cell. We plot the data
as a function of frequency for several wave vectors $\mathbf{q}$ traversing
along $\Gamma $--$M$ line of the BZ and for the largest diagonal elements of
the spin susceptibility that we find for Ni d$_{3z^{2}-r^{2}}$ orbitals. The
spin fluctuational spectrum is primarily located at small frequencies and
peaked around the energies $0.05eV$ or so. It exhibits significant momentum
dependence for Ni d$_{3z^{2}-r^{2}}$ orbitals which shows a broad peak
around the momenta (1/3,1/3,0)2$\pi /a$ but grows toward the edge of the BZ
at (1/2,1/2,0)2$\pi /a$. Another maximum was also found around the BZ point
(1/2,0,0)2$\pi /a.$ This indicates that the spin fluctuations are primarily
of antiferromagnetic character. Despite the long range antiferromagnetic
order is absent in La$_{3}$Ni$_{2}$O$_{7}$, this result is quite similar to
cuprates and iron based materials where doping is served to suppress the
long range magnetic order.

It is interesting to note that the range of $\omega _{sf}\symbol{126}50$ $%
meV $ is also seen for high--T$_{c}$ cuprates as peaks in imaginary spin
susceptibility accessible via the neutron scattering experiment\cite{INS}.
There is a famous 40 meV resonance which is visible in the superconducting
state\cite{40meV}. There are numerous angle resolved photoemission
experiments (ARPES) that show kinks in the one--electron spectra at the same
energy range\cite{ARPES-Kinks}. These kinks are sometimes interpreted as
caused by the electron--phonon interactions\cite{Lanzara}, but,
unfortunately, the calculated values of $\lambda _{e-p.}$ are known to be
small in the cuprates\cite{Savrasov-OKA,Louie}. Note also that for the
undoped antiferromagnetic cuprates, the spin wave spectra reside in the
energy range of 30 meV \cite{SpinWaves}.

We can judge about the values of $T_{c}$ using our estimated $\omega _{sf}%
\symbol{126}50$ $meV$ and the values of $\lambda _{eff}$ that we calculate
in Fig. \ref{FigLambdas}. For $\lambda _{eff}=0.1,$ the BCS$\ T_{c}=\omega
_{sf}\exp (-1/\lambda _{eff})\approx 0.02K$. Once we get closer to the SDW
instability, the effective coupling increases to the values 0.4 and the
corresponding BCS $T_{c}\approx 48K$. Given the exponential sensitivity of
the $T_{c},$ the latter value is certainly not far away from 80K range.

\section{IV\ Tight--Binding Calculations for La$_{3}$Ni$_{2}$O$_{7}$}

Several tight--binding calculations addressed the origin of
superconductivity in La$_{3}$Ni$_{2}$O$_{7}$ and the symmetry of the pairing
state in the recent literature \cite%
{Hu-arxiv,Qin-2023,Yang-2023,Tian-2023,Dagotto-arxiv,Lechermann-2023,SPM-arxiv,LongRange}%
. Although a few publications predicted the s$_{\pm }$ as leading pairing
symmetry \cite{Hu-arxiv,Qin-2023,Yang-2023,Tian-2023,Dagotto-arxiv}, results
based on diagonalizing the BCS gap equation have also emphasized the
appearance of the d$_{xy}$ pairing state \cite%
{Lechermann-2023,SPM-arxiv,LongRange}. 
\begin{figure}[tbp]
\includegraphics[height=0.373\textwidth,width=0.40\textwidth]{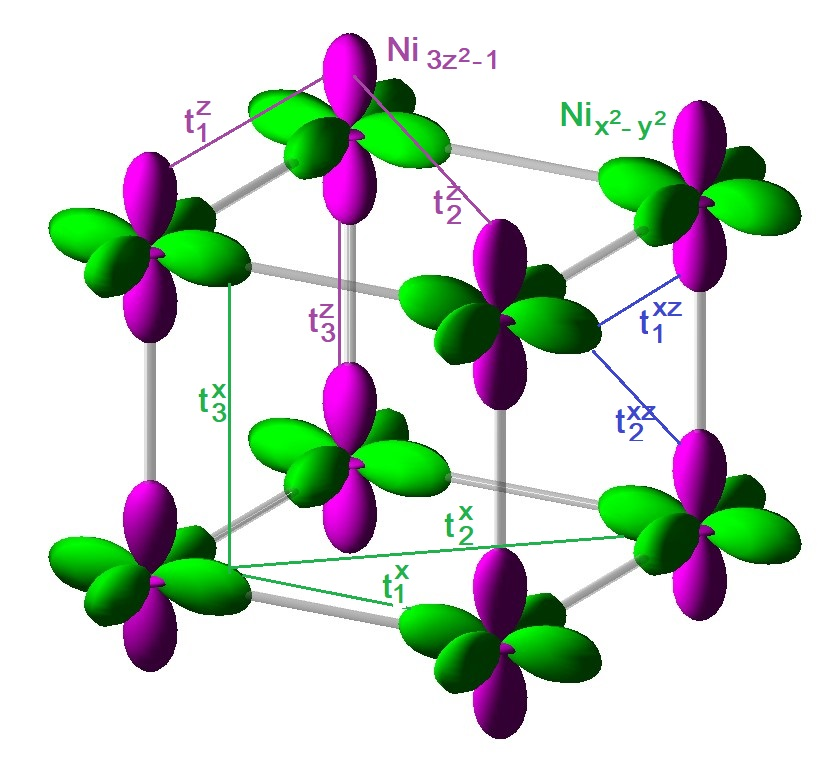}
\caption{Two--orbital ($x^{2}-y^{2}$ and $3z^{2}-r^{2}$) bilayer
tight--binding model employed for calculations of superconductivity in La$%
_{3}$Ni$_{2}$O$_{7}.$}
\label{FigTBModel}
\end{figure}

To validate these findings, we have performed our own tight--binding
simulations using the FLEX(RPA) method by utilizing the two--orbital (Ni-d$%
_{x^{2}-y^{2}}$ and Ni--d$_{3z^{2}-r^{2}})$ model that was proposed to
describe the electronic structure of La$_{3}$Ni$_{2}$O$_{7}$\cite{Bilayer1}$%
. $ We illustrate this model in Fig. \ref{FigTBModel}, whose parameters are
almost identical to those found in Ref.\cite{Bilayer1}: using shortcut
notations $x/z$ for Ni--$3d_{x^{2}-y^{2}}$ and Ni--$3d_{3z^{2}-r^{2}}$
orbitals, respectively, we have the on--site energy levels $\epsilon
_{x}=0.776$ eV$,\epsilon _{z}=0.409$ eV$,$ and the hopping integrals as
follows: $t_{1}^{x}=-0.483$ eV$,t_{2}^{x}=0.069$ eV$,t_{3}^{x}=0.005$ eV$%
,t_{1}^{z}=-0.11$ eV$,t_{2}^{z}=-0.017$ eV$,t_{3}^{z}=-0.335$ eV$%
,t_{1}^{xz}=0.239$ eV$,t_{2}^{xz}=-0.034$ eV$.$ (The only difference here is
the parameter $t_{3}^{z}$ which was quoted to be -0.635eV in Ref.\cite%
{Bilayer1}).

\begin{figure}[tbp]
\includegraphics[height=0.297\textwidth,width=0.40%
\textwidth]{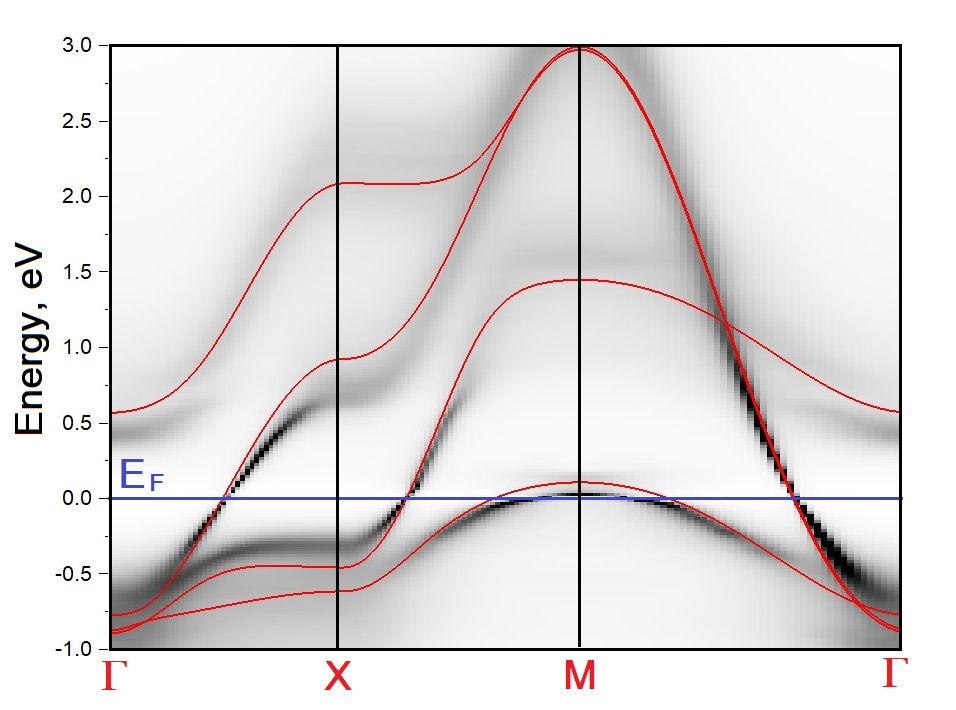}
\caption{Correlation effect using FLEX(RPA) method (black) on the electronic
structure of two orbital bilayer tight--binding model (red) for La$_{3}$Ni$%
_{2}$O$_{7}.$ }
\label{FigTBBands}
\end{figure}

Our tight--binding energy bands are illustrated in Fig.\ref{FigTBBands} by
red lines along the same high--symmetry lines as we used in Fig.\ref%
{FigBands}: We conclude that the fit bears close resemblance to the LDA
derived electronic structure. 

We further include the effect of Coulomb correlations using the FLEX(RPA)
method by calculating the self--energies and the single--particle Green
functions. We use the Coulomb interaction parameters as in Ref. \cite%
{SPM-arxiv}: the intraorbital $U$ is fixed to be 1.16 eV, and the exchange
integral $J$ is set to $U/4$ while the interorbital Coulomb integral $V=U-2J$%
. $\ $These parameters are smaller from the ones employed in full LDA+FLEX
calculation primarily due to the restricted Hilbert space at which the
tight--binding parameterization operates.

The result is illustrated in Fig.\ref{FigTBBands} where we plot the poles of
the Green function whose frequency dependence is shown by a varied gray
shading. In accord with the Fermi liquid theory, one can see that the
long--living quasiparticles (sharp black lines) are present in the vicinity
of the Fermi energy only, while the resonances become more diffusive as we
depart from the Fermi level. This is very similar to what we see in our full
LDA+FLEX calculation for the bands of predominantly Ni-d$_{x^{2}-y^{2}}$ and
Ni--d$_{3z^{2}-r^{2}}$ character, Fig \ref{FigBands}.

\begin{figure}[tbp]
\includegraphics[height=0.815\textwidth,width=0.40%
\textwidth]{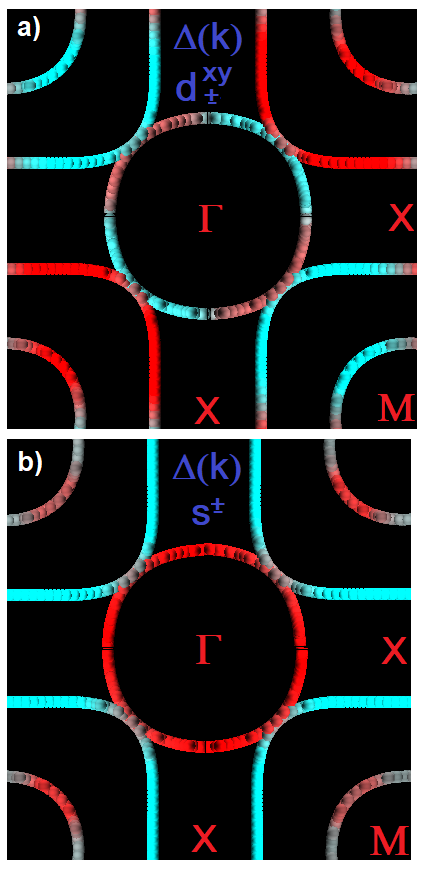}
\caption{{}Calculated superconducting energy gap $\Delta (\mathbf{k})$ for
singlet pairing in La$_{3}$Ni$_{2}$O$_{7}$ using numerical solution of the
linearized BCS gap equation with the pairing interaction evaluated using
tight--binding parameterization of the LDA energy bands and FLEX(RPA)
method. Blue/red color marks points at the two--dimensional Fermi surface
(viewed from the top) that corresponds to the negative/positive values of $%
\Delta (\mathbf{k}),$ while the values around zero are colored in gray. Case
a) is the eigenstate with d$_{xy}$ symmetry and b) is the eigenstate with s$%
_{\pm }$ symmetry.}
\label{FigTBGaps}
\end{figure}

We further repeat the exact diagonalization of the BCS\ gap equation using
the RPA\ pairing interaction calculated within the two--orbital bilayer
model. The behavior of $\Delta _{S}(\mathbf{k}j)$ for the two highest
eigenvalues is illustrated in Fig.\ref{FigTBGaps} where the plot \ref%
{FigTBGaps}(a) shows the behavior characteristic for the d$_{xy}$ symmetry
while the plot \ref{FigTBGaps}(b) shows the gap function of the s$_{\pm }$
symmetry. One can see that momentum dependence of the both gap functions is
very similar to the ones that we calculate using full LDA+FLEX method, Fig. %
\ref{FigGaps}.

\begin{figure}[tbp]
\includegraphics[height=0.338\textwidth,width=0.40%
\textwidth]{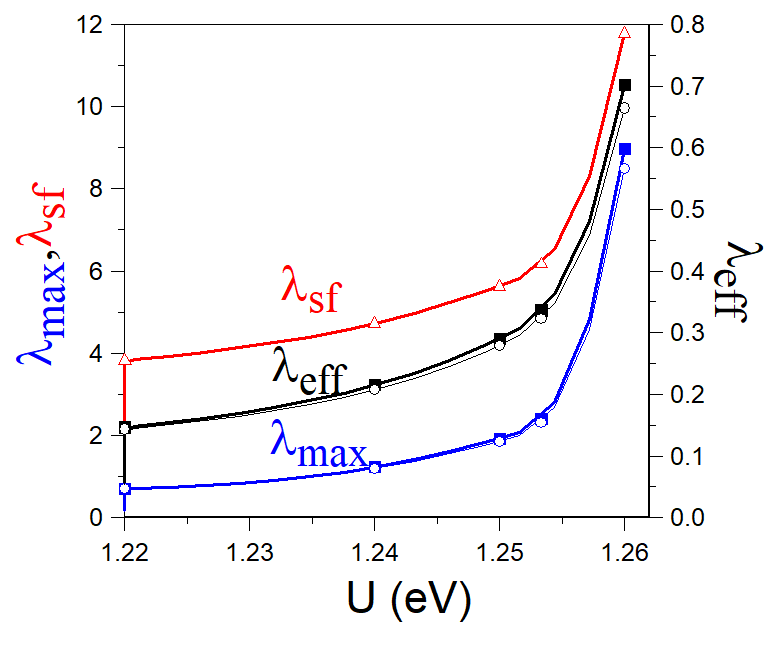}
\caption{{}Calculations using tight--binding parameterization of the LDA\
energy bands and the FLEX(RPA)\ method of the spin fluctuational mass
enhancement, $\protect\lambda _{sf},$ and of the two highest eiogenvalues $%
\protect\lambda _{\max }$ corresponding to s$_{\pm }$ (squares) and d$_{xy}$
(circles) symmetries of the linearized BCS equation as a function of the
on--site Hubbard interaction U for the two--orbital bilayer model of La$_{3}$%
Ni$_{2}$O$_{7}$. The effective coupling constant $\protect\lambda _{eff}=%
\protect\lambda _{\max }/(1+\protect\lambda _{sf})$ is shown on the scale
given on the right. }
\label{FigTBLambdas}
\end{figure}

It is however interesting to note that in the case of the tight--binding
calculation, the solution of the s$_{\pm }$ symmetry comes out always
slightly more favorable than the solution of the d$_{xy}$ symmetry. We
illustrate this result by plotting the eigenvalue $\lambda _{\max }$ as a
function of the parameter $U$ on Fig. \ref{FigTBLambdas}, where the highest
eigenstate marked by squares corresponds to the s$_{\pm }$ symmetry while
the next highest marked by circles is of the d$_{xy}$ symmetry. 

Our tight--binding FLEX(RPA) calculation is found in complete agreement with
the similar calculation that has recently appeared in the literature \cite%
{SPM-arxiv}. However it is somewhat different from our LDA+FLEX calculation
shown \ref{FigLambdas} where we found that d$_{xy}$ symmetry is more
favorable for the range of $U$'s less than 3.2 eV. It is clear that the
inclusion of the full LDA energy bands and the wave functions can be a
possible source of this discrepancy.

We further calculate the mass enhancement parameter $\lambda _{sf}$ due to
spin fluctuations as a function of $U$. The result is shown in Fig. \ref%
{FigTBLambdas} where one can see that the mass enhancement raises rapidly
once the system is tuned to the SDW\ instability. The same is seen in the
dependence of the effective coupling constant $\lambda _{eff}=\lambda _{\max
}/(1+\lambda _{sf})$ which is illustrated on the right scale of Fig.\ref%
{FigTBLambdas}. The $\lambda _{eff}$ can reach pretty high values \symbol{126%
}0.7 as we vary $U$ and the corresponding BCS\ T$_{c}=\omega _{sf}\exp
(-1/\lambda _{eff})\approx 139K$ using $\omega _{sf}=50$ $meV.$ This
estimate would however assume very large values of $\lambda _{sf}\approx 12.$
More modest values of $\lambda _{sf}\approx 5$ correspond to $\lambda
_{eff}\approx 0.3$ for which the BCS T$_{c}$ is estimated to be about $20K.$ 

Clearly, one cannot expect the high accuracy from the RPA regarding the
precise determination of the pairing interaction and the corresponding
extraction of the T$_{c}$ but one can judge that the inclusion of the full
LDA derived band structures lowers the estimate of the effective coupling
constant as compared to the tight--binding calculation. This trend was also
found by us in our recent exploration of the cuprate superconductor HgBa$%
_{2} $CuO$_{4}$\cite{Hg-FLEX}.

\section{V. Conclusion}

In conclusion, we used recently developed LDA+FLEX method to study spin
fluctuational mechanism of superconductivity in recently discovered La$_{3}$%
Ni$_{2}$O$_{7}$ under pressure. Based on this procedure, the superconducting
scattering matrix elements between the Cooper pairs have been evaluated
numerically, and the linearized BCS\ gap equation was exactly diagonalized.
Our main result is the competition between d$_{xy}$ and s$_{\pm }$ pairing
symmetries for the most favorable eigenstate of the superconducting order
parameter, which unfortunately cannot be precisely determined given the
sensitivity of the result to the values of Hubbard parameter $U$ used in the
simulation. Since both symmetries can be distinguished by the
presence/absence of nodes in the gap function, it should be straightforward
to sort this out with currently on--going experiments.

The superconducting coupling constant $\lambda _{\max }$ as the highest
eigenvalue of the BCS\ gap equations has been extracted together with the
spin fluctuational mass enhancement $m^{\ast }/m_{LDA}=1+\lambda _{sf}.$ The
effective coupling constant $\lambda _{eff}=\lambda _{\max }/$($1+\lambda
_{sf})$ was deduced as a function of $U$, but found to be modest and
incapable to deliver high values of $T_{c}$ unless $U$ is tuned to be close
to SDW. We have also performed tight--binding calculations using recently
proposed two--orbital bilayer model which confirmed our conclusions.

\end{document}